\documentclass[fleqn,10pt]{SelfArx} 

\definecolor{color1}{RGB}{0,0,90} 
\definecolor{color2}{RGB}{0,20,20} 

\usepackage{hyperref} 
\hypersetup{hidelinks,colorlinks,breaklinks=true,urlcolor=color2,citecolor=color1,linkcolor=color1,bookmarksopen=false,pdftitle={Title},pdfauthor={Author},urlcolor=blue}
\usepackage{graphicx}
\usepackage{natbib}
\usepackage{amsmath}
\usepackage{amsfonts}
\usepackage{multirow}
\usepackage{subfigure}
\usepackage{tikz}
\usetikzlibrary{calc}

\DeclareMathOperator{\sgn}{sgn}
\DeclareMathOperator{\atanh}{atanh}
\DeclareMathOperator{\atan}{atan}

\usepackage{mathtools,amssymb,lipsum}

\usepackage{cuted}
\usepackage{boondox-calo}

\newcommand{\n}[1]{\mathrm{#1}}
\newcommand{\parde}[2]{\frac{\partial#1}{\partial#2}}
\newcommand{\ud}{\mathrm{d}}
\newcommand{\xhat}{\mathbf{\hat{x}}}
\newcommand{\yhat}{\mathbf{\hat{y}}}
\newcommand{\zhat}{\mathbf{\hat{z}}}
\newcommand{\nhat}{\mathbf{\hat{n}}}
\newcommand{\rhat}{\mathbf{\hat{r}}}
\newcommand{\thetahat}{\mathbf{\hat{\theta}}}

\newcommand{\dDdx}{\parde{D}{x}}
\newcommand{\dDdy}{\parde{D}{y}}
\newcommand{\dDdz}{\parde{D}{z}}

\JournalInfo{Published in Journal of Magnetism and Magnetic Materials, Vol. 507, 166799, 2020} 
\Archive{\href{https://doi.org/10.1016/j.jmmm.2020.166799}{DOI: 10.1016/j.jmmm.2020.166799}} 

\PaperTitle{The magnetic field from a homogeneously magnetized cylindrical tile} 

\Authors{K. K. Nielsen and R. Bj\o{}rk} 
\affiliation{\textit{Department of Energy Conversion and Storage, Technical University of Denmark - DTU, Anker Engelunds Vej 1, DK-2800 Kgs. Lyngby, Denmark}} 
\affiliation{*\textbf{Corresponding author}: rabj@dtu.dk} 

\newcommand{\keywordname}{Keywords} 

\Abstract{The magnetic field of a homogeneously magnetized cylindrical tile geometry, i.e. an angular section of a finite hollow cylinder, is found. The field is expressed as the product between a tensor field describing the geometrical part of the problem and a column vector holding the magnetization of the tile. Outside the tile, the tensor is identical to the demagnetization tensor. We find that four components of the tensor, $N_{xy},N_{xz},N_{yz}$ and $N_{zy}$, can be expressed fully analytically, while the five remaining components, $N_{xx},N_{yx},N_{yy},N_{zx}$ and $N_{zz}$, contain integrals that have to be evaluated numerically. When evaluated numerically the tensor is symmetric. A comparison between the found solution, implemented in the open source magnetic framework MagTense, and a finite element calculation of the magnetic flux density of a cylindrical tile shows excellent agreement.}

\begin{document}

\flushbottom 

\maketitle 


\thispagestyle{empty} 

\section{Introduction}
A magnetic field is used in a very large variety of applications. Calculating the magnetic field generated by a homogeneously magnetized magnet of a specific shape and size is relevant for several applications, including permanent magnet motors and generators, magnetic bearings and nuclear magnetic resonance (NMR) systems. The magnetic field generated by a permanent magnet or a system of these is often calculated using a finite element framework (FEM) approach, which is limited in precision and speed by the fineness of the finite element mesh used in the calculation. However, the magnetic field generated by a few specific magnet geometries can be expressed analytically if the magnet is uniformly magnetized. This is for example the case for an ellipsoid or a cylinder of infinite length\cite{Osborn_1945}. The magnetic field is often calculated through the use of the demagnetization tensor, which expresses the relation between the magnetic field and the magnetization
\begin{eqnarray}\label{Eq.Demag_def}
\mathbf{H}=\mathbf{H}_\textrm{appl}-\mathbb{N}_\n{d}\cdot{}\mathbf{M}
\end{eqnarray}
where $\mathbf{H}$ is the magnetic field, $\mathbf{H}_\textrm{appl}$ is an externally applied field, $\mathbf{M}$ is the magnetization and $\mathbb{N}_\n{d}$ is the demagnetization tensor field, which is a rank-2 symmetric tensor \cite{Moskowitz_1966}.

The magnetic field anywhere in space can be calculated analytically through the demagnetization tensor field for a few specific geometries with uniform magnetization, e.g. for a rectangular prism \cite{Joseph_1965,Smith_2010} or a hollow sphere \cite{Prat-Camps2016}. If analytical expressions for the tensor are not known, the internal field within the magnet can be estimated using an average demagnetization factor \cite{Aharoni_1998,Joseph_1966}, which can also be computed for non-solid magnetic samples, i.e. powder samples \cite{Breit_1922,Bleaney_1941,Bjoerk_2013,Arzbacher_2015}.

In this work we consider the magnetic field generated by a solid cylindrical tile. The cylindrical tile is an angular section of a hollow finite cylinder, as shown in Fig. \ref{Fig_Cylinder_tile}. This geometry is important in several types of applications, for example as segments of a permanent magnet magnetic resonance imaging (MRI) scanner, as magnetic pieces of a generator/motor or as segmented pieces of a Halbach cylinder. For the cylindrical tile magnet geometry there has been a previous attempt to calculate the magnetic field \cite{Ravaud_2009}, but no closed-form analytical solution could be obtained for any of the components. The derived expressions for the components of the magnetic field all included definite integrals that have to be evaluated numerically. The cylindrical tile geometry can be seen as a continuation of a previous work, which calculated the magnetic field from a 3D permanent magnetic ring \cite{Ravaud_2008}. The magnetic field generated by a whole cylinder is also known \cite{Kraus_1973,Chen_1991,Chen_2006,Caciagli_2018}. The magnetic field generated by the cylindrical tile has also been considered in the less complex case of a cylinder of infinite length, i.e. a two dimensional system. In two dimensions, the magnetic field produced by a cylindrical tile has been studied extensively, and the field has been calculated numerous times, with focus on different applications such as magnetic gears \cite{Lubin_2010}, motors with Halbach arrays \cite{Markovic_2009}, segmented two dimensional Halbach cylinder \cite{Shi_2012} and brushless permanent magnet motors \cite{Zhu_1993}.

In this work we will derive analytical expressions of the components of the magnetic flux density formulated as a tensor field for a cylindrical tile, for as many of the components as possible. The magnetic field can then easily be computed from the magnetic flux density and the magnetization. The computer code of this model (written in Fortran with a Matlab interface) is publicly available as a part of the MagTense code \cite{MagTense}.

\section{The magnetic field of a cylindrical tile}
\begin{figure*}[!t]
\begin{tikzpicture}

	\coordinate (O) at (0.5,0.5);
	\coordinate (x1) at (4,1.6569);
	\coordinate(x2) at (7,2.8995);
	\coordinate(x3) at (3.0615,3.0615);
	\coordinate(x4) at (5.3576,5.3576);
	\def\rone{4.3296}
	\def\rtwo{7.5767}
	
	\draw[semithick,->] (0,0) -- (8,0);
	\draw[semithick,->] (0,0) -- (0,6);
	\draw[semithick,->] (0,0) -- (-2,-1);
	
	\draw (7.5,-0.5) node {$x$};
	\draw (-0.5,6) node {$y$};
	\draw (-1.5,-1.2) node {$z$};
	
	\draw[semithick] (x1) -- (x2);
	\draw[dashed,color=gray] ($(x1)+(O)$) -- ($(x2)+(O)$);
	
	\draw[semithick] (x1) arc (22.5:45:\rone);
	\draw[dashed,color=gray] ($(O)+(x1)$) arc (22.5:45:\rone);
	
	\draw[semithick] (x2) arc (22.5:45:\rtwo);
	\draw[semithick] ($(x2)+(O)$) arc (22.5:45:\rtwo);
	
	\draw[semithick] (x3) -- (x4);
	\draw[semithick] ($(x3)+(O)$) -- ($(x4)+(O)$);
	
	\draw[dashed,color=gray] (x1) -- ($(x1)+(O)$);
	
	\draw[semithick] (x2) -- ($(x2)+(O)$);
	
	\draw[dashed,color=gray] (0,0) -- (x1);
	\draw[dashed,color=gray] (0,0) -- (x3);
	
	\draw[dashed,color=gray] (2,0) arc (0:22.5:2);
	
	\draw[dashed,color=gray] (3,0) arc (0:45:3);
	
	\draw (2.3,0.5) node {$\theta_1$};
	\draw (2.9,1.8) node {$\theta_2$};
	
	\draw ($(x1)+(-0.5,0.5)$) node {$r_1$};
	\draw ($(x2)+(-1,1)$) node {$r_2$};
	
	\draw ($(x2)+(0,-0.5)$) node {$z_1$};
	\draw ($(x2)+(O)+(0.3,-0.3)$) node {$z_2$};

    \draw ($(x2)+(0.1,2.2)$) node {$Arc$};
    \draw ($(x2)+(0.6,1.7)$) node {$surface$};
    \draw ($(x2)+(-2.0,0.6)$) node {$Horizontal$};
    \draw ($(x2)+(-2.0,0.1)$) node {$surface$};
    \draw ($(x2)+(-3.4,2.0)$) node {$Vertical$};
    \draw ($(x2)+(-3.4,1.5)$) node {$surface$};
\end{tikzpicture}
	\caption{The cylindrical tile considered in this work.}
    \label{Fig_Cylinder_tile}
    \end{figure*}
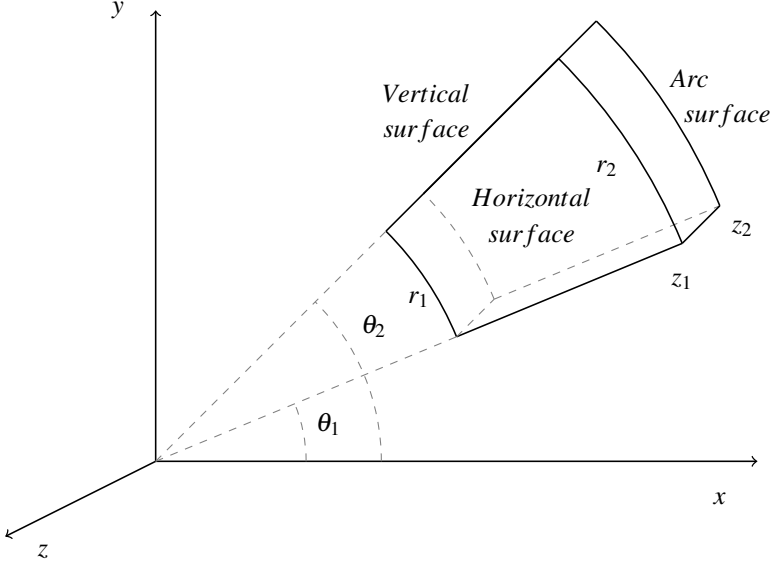

We consider a cylindrical tile in 3D composed of six surfaces. The cylindrical tile is completely specified by six parameters, namely the angular span of the tile from $\theta_1$ to $\theta_2$, the height of the cylinder from $z_1$ to $z_2$, and the internal, $r_\n{1}$, and external radii of the cylinder, $r_\n{2}$, as illustrated in Fig. \ref{Fig_Cylinder_tile}. We assume the cylinder tile to be homogeneously magnetized with the magnetization $\mathbf{M}=(M_x,M_y, M_z)$. Our objective is then to find the magnetic flux density, $\mathbf{B}$, at any point $\mathbcal{r}$, be it inside or outside the cylinder tile. Once the magnetic flux density is known, the magnetic field can easily be calculated from $\mathbf{B} = \mu{}_{0}(\mathbf{H} + \mathbf{M})$, where the vacuum permeability is denoted $\mu_0$.

We are only interested in the magnetostatic limit and so no free volume currents are flowing. Following \cite{Griffiths} we may write the magnetic vector potential assuming the Coulomb gauge, $\nabla{}\cdot{}\mathbf{A} = 0$, at the position $\mathbcal{r}$ as:
\begin{equation}
\mathbf{A}(\mathbcal{r})=\frac{\mu_0}{4\pi}\int_{V'}\frac{\nabla'\times\mathbf{M}(\mathbcal{r}')}{|\mathbcal{r}-\mathbcal{r}'|}\ud V' + \frac{\mu_0}{4\pi}\int_S\frac{\mathbf{M}(\mathbcal{r}')\times\nhat(\mathbcal{r}')}{|\mathbcal{r}-\mathbcal{r}'|}\ud a'.
\end{equation}
Note that there are two sets of coordinates. The coordinates marked with a $'$ are the coordinates of the structure that creates the magnetic field, whereas the non-marked coordinates are to the point at which the field is evaluated. It is seen that the first term in the equation is the zero-vector since the magnetization is assumed constant. This also means that the marked coordinates thus refer to the surface of the object that creates the magnetic field. The unit vector normal to the surface with surface element $\ud a'$ is $\nhat$.

Given $\mathbf{B}=\nabla\times\mathbf{A}$ we now have:
\begin{equation}
\mathbf{B}(\mathbcal{r})=\frac{\mu_0}{4\pi}\nabla\times\int_S\frac{\mathbf{M}(\mathbcal{r}')\times\nhat(\mathbcal{r}')}{|\mathbcal{r}-\mathbcal{r}'|}\ud a'.
\end{equation}
Note that the curl-operator is in non-marked coordinates. As the curl-operator and the integration are with respect to different sets of coordinates, the order of these are interchangable.

We now consider cylindrical coordinates, where the unit vectors can be expressed in the Cartesian base as:
\begin{eqnarray}
\nhat_1&=&\rhat=(\cos\theta',\sin\theta',0)\nonumber\\
\nhat_2&=&\thetahat=(-\sin\theta',\cos\theta',0)\nonumber\\
\nhat_3&=&\zhat=(0,0,1),
\end{eqnarray}
noting that they are given in marked coordinates. Note however that the base coordinate system is the same regardless of whether marked coordinates are used or not, and therefore the mark on e.g. $\rhat$ or on $\xhat$ will be omitted.

With these the integral can be expressed as
\begin{equation}
\mathbf{B}(\mathbcal{r})=\frac{\mu_0}{4\pi}\nabla\times\int_{S}D\left(\mathbcal{r}-\mathbcal{r}'\right)\mathbf{M}(\mathbcal{r}')\times\nhat(\mathbcal{r}')\ud a'.
\end{equation}

with
\begin{eqnarray}
D&=&\frac{1}{\left|\mathbcal{r}-\mathbcal{r}'\right|}=\frac{1}{\sqrt{(x-r'\cos\theta')^2+(y-r'\sin\theta')^2+(z-z')^2}}\nonumber\\
r'&=&\sqrt{x'^2+y'^2}
\end{eqnarray}

The components of the magnetic flux density are computed by integrating along the entire closed surface of the cylindrical tile. Here the six surfaces of the cylindrical tile are most easily integrated in cylindrical coordinates. In this orthonormal basis, the cylindrical tile is fully symmetric with respect to all axes, which reduces the number of integrals to three.

\subsection{Surface 1 - the arc surface}
We first consider the surface for which $r'=r_s$ is constant, i.e. the arc surface. Here $r_s$ attains the values $r_1$ or $r_2$. We consider the surface for which the normal points outwards, i.e. the outer arc surface for which $r_s=r_2$. In order to find the flux density we need to integrate the un-marked curl of the surface current density:
\begin{eqnarray}
&&\nabla\times \left(D\left(-\xhat M_z\sin\theta'+\yhat M_z\cos\theta'+\zhat(M_x\sin\theta'-M_y\cos\theta')\right)\right)\nonumber\\
&=&\xhat\left((M_x\sin\theta'-M_y\cos\theta')\dDdy-M_z\cos\theta'\dDdz\right)\nonumber\\
&-&\yhat\left((M_x\sin\theta'-M_y\cos\theta')\dDdx+M_z\sin\theta'\dDdz\right)\nonumber\\
&+&\zhat\left(M_z\cos\theta'\dDdx+M_z\sin\theta'\dDdy\right),
\end{eqnarray}

The flux density for surface 1 then becomes:
\begin{eqnarray}
B_x^1(\mathbcal{r})&=&\frac{\mu_0r_2}{4\pi}\int_{z_1}^{z_2}\int_{\theta_1}^{\theta_2}\left.(M_x\sin\theta'-M_y\cos\theta')\dDdy\right.\nonumber\\
&&\left.-M_z\cos\theta'\dDdz\ud z'\ud\theta'\right|_{r'=r_2}\nonumber\\
B_y^1(\mathbcal{r})&=&-\frac{\mu_0r_2}{4\pi}\int_{z_1}^{z_2}\int_{\theta_1}^{\theta_2}\left.(M_x\sin\theta'-M_y\cos\theta')\dDdx\right.\nonumber\\
&&\left.+M_z\sin\theta'\dDdz\ud z'\ud\theta'\right|_{r'=r_2}\nonumber\\
B_z^1(\mathbcal{r})&=&\frac{\mu_0r_2}{4\pi}\int_{z_1}^{z_2}\int_{\theta_1}^{\theta_2}M_z\left.\left(\cos\theta'\dDdx+\sin\theta'\dDdy\right)\ud z'\ud\theta'\right|_{r'=r_2}.
\end{eqnarray}
For the inner arc surface, the normal points inwards and a sign change is required as well as changing where $r'$ is evaluated from $r_2$ to $r_1$.

\subsection{Surface 2 - the vertical surface}
The second surface is defined for constant $\theta'$ and thus lies in the $r'z'$-plane. We consider the surface normal along the positive azimuthal unit-vector, $\thetahat$, and obtain:
\begin{eqnarray}
&&\nabla\times(D(-\xhat M_z\cos\theta' -\yhat M_z\sin\theta' \nonumber\\
&&+\zhat(M_x\cos\theta'+M_y\sin\theta')))\nonumber\\
&=&\xhat\left((M_x\cos\theta'+M_y\sin\theta')\dDdy+M_z\sin\theta'\dDdz\right)\nonumber\\
&-&\yhat\left((M_x\cos\theta'+M_y\sin\theta')\dDdx+M_z\cos\theta'\dDdz\right)\nonumber\\
&+&\zhat\left(M_z\cos\theta'\dDdy-M_z\sin\theta'\dDdx\right),
\end{eqnarray}
which gives the following contribution to the flux density:
\begin{eqnarray}
B_x^2(\mathbcal{r})&=&\frac{\mu_0}{4\pi}\int_{r_1}^{r_2}\int_{z_1}^{z_2}\left.(M_x\cos\theta'+M_y\sin\theta')\dDdy\right.\nonumber\\
&&\left.+M_z\sin\theta'\dDdz\ud r'\ud z'\right|_{\theta'=\theta_2}\nonumber\\
B_y^2(\mathbcal{r})&=&-\frac{\mu_0}{4\pi}\int_{r_1}^{r_2}\int_{z_1}^{z_2}\left.(M_x\cos\theta'+M_y\sin\theta')\dDdx\right.\nonumber\\
&&\left.+M_z\cos\theta'\dDdz\ud r'\ud z'\right|_{\theta'=\theta_2}\nonumber\\
B_z^2(\mathbcal{r})&=&\frac{\mu_0}{4\pi}\int_{r_1}^{r_2}\int_{z_1}^{z_2}M_z\left.\left(\cos\theta'\dDdy\right.\right.\nonumber\\
&&\left.\left.-\sin\theta'\dDdx\right)\ud r'\ud z'\right|_{\theta'=\theta_2}\nonumber
\end{eqnarray}
The surface with the normal anti-parallel to the azimuthal unit-vector is found by a change of sign on the flux density and replacing where $\theta'$ is evaluated from $\theta_2$ to $\theta_1$.

\subsection{Surface 3 - the horizontal surface}
Surface 3 is defined for constant $z'$ and lies in the $r'\theta'$-plane. Considering the normal parallel to the positive $z-$direction, i.e. $\zhat$ we get:
\begin{eqnarray}
&&\nabla\times(D(\xhat M_y-\yhat M_x))\nonumber\\
&=&\xhat M_x\dDdz\nonumber+\yhat M_y\dDdz-\zhat\left(M_x\dDdx+M_y\dDdy\right),
\end{eqnarray}
leading to the following:
\begin{eqnarray}
B_x^3(\mathbcal{r})&=&\frac{\mu_0}{4\pi}\int_{r_1}^{r_2}\int_{\theta_1}^{\theta_2}\left.M_x\dDdz{}r' \ud r'\ud\theta'\right|_{z'=z_2}\nonumber\\
B_y^3(\mathbcal{r})&=&\frac{\mu_0}{4\pi}\int_{r_1}^{r_2}\int_{\theta_1}^{\theta_2}\left.M_y\dDdz{}r' \ud r'\ud\theta'\right|_{z'=z_2}\nonumber\\
B_z^3(\mathbcal{r})&=&-\frac{\mu_0}{4\pi}\int_{r_1}^{r_2}\int_{\theta_1}^{\theta_2}\left.\left(M_x\dDdx+M_y\dDdy{}\right)r'\ud r'\ud\theta'\right|_{z'=z_2}\nonumber\\
\end{eqnarray}
The surface with its normal vector anti-parallel to the $z-$direction requires a change of sign and replacing where $z'$ is evaluated from $z_2$ to $z_1$.

\subsection{Tensor components}
From the above calculations it is seen that the local flux density at the point $\mathbcal{r}$ from a homogeneously magnetized cylindrical tile can be written as the product between a tensor and the magnetization vector:
\begin{eqnarray}
\mathbf{B}(\mathbcal{r})=\frac{\mu_0}{4\pi}\mathbb{N}(\mathbcal{r}-\mathbcal{r}_p)\cdot\mathbf{M}(\mathbcal{r}_p),
\end{eqnarray}
We explicitly remark that the magnetization vector has been assumed to be constant. The center of the tile is denoted $\mathbcal{r}_p=(x_p,y_p,z_p)$.

Outside the tile, the magnetic field is equal to the magnetic flux density given above with a factor of $\mu_0$. Therefore the tensor $\mathbb{N}$ is identical, with a factor of $-1/4\pi$, to the demagnetization tensor defined in Eq. \ref{Eq.Demag_def} outside the tile. Within the tile, the demagnetization tensor as defined in Eq. \ref{Eq.Demag_def} and the magnetic field can be obtained from the relation $\mathbf{B}=\mu_0(\mathbf{H}+\mathbf{M})$, as the above expression give $\mathbf{B}$ both within and outside the tile.

The total flux density may be found by summing the contributions from all the surfaces of the cylindrical tile as given above. We introduce the following nomenclature for simplicity:
\begin{eqnarray}
\mathcal{A}(r') &=& \int_{z_1}^{z_2}\int_{\theta_1}^{\theta_2}\sin\theta'\dDdy\ud z'\ud\theta' \nonumber\\
\mathcal{B}(\theta{}') &=& \int_{r_1}^{r_2}\int_{z_1}^{z_2}\dDdy\ud r'\ud z' \nonumber\\
\mathcal{C}(z') &=& \int_{r_1}^{r_2}\int_{\theta_1}^{\theta_2}\dDdz r'\ud r'\ud\theta' \nonumber\\
\mathcal{D}(r') &=& \int_{z_1}^{z_2}\int_{\theta_1}^{\theta_2}\cos\theta'\dDdy\ud z'\ud\theta' \nonumber\\
\mathcal{E}(r') &=& \int_{z_1}^{z_2}\int_{\theta_1}^{\theta_2}\cos\theta'\dDdz\ud z'\ud\theta' \nonumber\\
\mathcal{F}(\theta{}') &=& \int_{r_1}^{r_2}\int_{z_1}^{z_2}\dDdz\ud r'\ud z' \nonumber\\
\mathcal{G}(r') &=& \int_{z_1}^{z_2}\int_{\theta_1}^{\theta_2}\sin\theta'\dDdx\ud z'\ud\theta' \nonumber\\
\mathcal{H}(\theta{}') &=& \int_{r_1}^{r_2}\int_{z_1}^{z_2}\dDdx\ud r'\ud z' \nonumber\\
\mathcal{I}(r') &=& \int_{z_1}^{z_2}\int_{\theta_1}^{\theta_2}\cos\theta'\dDdx\ud z'\ud\theta'\nonumber\\
\mathcal{J}(r') &=& \int_{z_1}^{z_2}\int_{\theta_1}^{\theta_2}\sin\theta'\dDdz\ud z'\ud\theta'\nonumber\\
\mathcal{K}(z') &=& \int_{r_1}^{r_2}\int_{\theta_1}^{\theta_2}\dDdx{}r'\ud r'\ud\theta'\nonumber\\
\mathcal{L}(z') &=& \int_{r_1}^{r_2}\int_{\theta_1}^{\theta_2}\dDdy{}r'\ud r'\ud\theta'\nonumber
\end{eqnarray}

The integrals above are all functions of the unmarked coordinates $(x,y,z)$ as well as a constant marked coordinate. In order to shorten the notation in the following, the integrals are given as a function of the constant marked coordinate only.

Using the integral expressions, the components of the tensor, $\mathbb{N}$, are given by
\begin{eqnarray}
N_{xx}&=&r_2\mathcal{A}(r_2)-r_1\mathcal{A}(r_1)+\cos\theta_2\mathcal{B}(\theta{}_2)-\cos\theta_1\mathcal{B}(\theta{}_1)\nonumber\\
&&+\mathcal{C}(z_2)-\mathcal{C}(z_1)\nonumber\\
N_{xy}&=&r_1\mathcal{D}(r_1)-r_2\mathcal{D}(r_2)+\sin\theta_2\mathcal{B}(\theta{}_2)-\sin\theta_1\mathcal{B}(\theta{}_1)\nonumber\\
N_{xz}&=&r_1\mathcal{E}(r_1)-r_2\mathcal{E}(r_2)+\sin\theta_2\mathcal{F}(\theta_2)-\sin\theta_1\mathcal{F}(\theta_1)\nonumber\\
N_{yx}&=&r_1\mathcal{G}(r_1)-r_2\mathcal{G}(r_2)+\cos\theta_1\mathcal{H}(\theta_1)-\cos\theta_2\mathcal{H}(\theta_2)\nonumber\\
N_{yy}&=&r_2\mathcal{I}(r_2)-r_1\mathcal{I}(r_1)+\sin\theta_1\mathcal{H}(\theta_1)-\sin\theta_2\mathcal{H}(\theta_2)\nonumber\\
&&+\mathcal{C}(z_2)-\mathcal{C}(z_1)\nonumber\\
N_{yz}&=&r_1\mathcal{J}(r_1)-r_2\mathcal{J}(r_2)+\cos\theta_1\mathcal{F}(\theta_1)-\cos\theta_2\mathcal{F}(\theta_2)\nonumber\\
N_{zx}&=&\mathcal{K}(z_1)-\mathcal{K}(z_2)\nonumber\\
N_{zy}&=&\mathcal{L}(z_1)-\mathcal{L}(z_2)\nonumber\\
N_{zz}&=&r_2\mathcal{A}(r_2)-r_1\mathcal{A}(r_1)+r_2\mathcal{I}(r_2)-r_1\mathcal{I}(r_1)+\cos\theta_2\mathcal{B}(\theta_2)\nonumber\\
&&-\cos\theta_1\mathcal{B}(\theta_1)+\sin\theta_1\mathcal{H}(\theta_1)-\sin\theta_2\mathcal{H}(\theta_2)
\end{eqnarray}

\subsection{Evaluating the integrals}
The integrals given above can more easily be evaluated by introducing a rotation and translation trick following \cite{Varga1998}. The goal is to express the integrals as a function of $x$ and the primed coordinates only, thus letting $y=z=0$.

Along the $z-$axis this is achieved by changing the integration limits for all integrals over $\ud z'$ by subtraction of the value of the $z-$coordinate. For the $y$-coordinate we rotate the point of interest about the $z-$axis, so that it lies on the $x$-axis. The rotation is defined by the angle $\psi=\tan^{-1}\left(\frac{y}{x}\right)$. The integration limits of the integrals over $\ud\theta'$ are changed by subtracting this angle. Finally, the magnetic flux density should be rotated back about the $z-$axis with the angle $-\psi$.

The evaluated expressions for the components of the tensor field are given below. Some integrals have been evaluated using Rubi, the Rule-based Integrator \cite{Rich2018}. The integral expressions of the components of the tensor field can most easily be written by introducing a few helper-functions. These are defined as follows:
\begin{eqnarray}
\mathbb{A}(r,x,\theta,z)&=&\sqrt{r^2-2xr\cos\theta+x^2+z^2}\nonumber\\
\mathbb{B}(x,\theta,z)&=&x^2(\cos^2\theta-1)-z^2\nonumber\\
\mathbb{C}(r,x)&=&\frac{4rx}{(r+x)^2}\nonumber\\
\mathbb{D}(\theta)&=&\cos\frac{\theta}{2}\nonumber\\
\mathbb{E}(r,x,z)&=&2\sqrt{\frac{rx}{(r+x)^2+z^2}}\nonumber\\
\mathbb{F}_\pm&=&\frac{\mathbb{A}(r,x,\theta,z_s)\pm r}{\sqrt{x^2+z_s^2}}\nonumber
\end{eqnarray}

The integral expressions of the components of the tensor field given below are given as indefinite integrals to compact the notation. As mentioned previously, the point of interest is defined as $\mathbf{r}=(x,0,0)$. Thus, the appropriate integration limits should be inserted in $(r',\theta',z')$, respectively. Subscript $s$ indicates a constant value on a surface, e.g. $r_s$ is the value of $r'$ on a $\theta' z'$ surface, i.e. $r_1$ or $r_2$. Incomplete elliptic integrals of the first, second and third kind are denoted $F$, $E$ and $\Pi$, respectively. Using these functions, the integrals can be expressed as given below, respecting the singularities in the solutions given in the appendix.
\begin{strip}
  \begin{align}
\mathcal{A}(r_s,\theta{}',z') &=& \frac{1}{r_s}\frac{1}{2}\frac{z'}{x^2\sqrt{(r_s+x)^2+z'^2}}\sgn\left(\sin\frac{\theta'}{2}\right)\bigg((r_s-x)^2\Pi\left\{\mathbb{D}(\theta'),\mathbb{C}(r_s,x),\mathbb{E}(r_s,x,z')\right\}\nonumber\\
&&+((r_s+x)^2+z'^2)E\left\{\mathbb{D}(\theta'),\mathbb{E}(r_s,x,z')\right\}-2(r^2+x^2+\frac{1}{2}z'^2)F\left\{\mathbb{D}(\theta'),\mathbb{E}(r_s,x,z')\right\}\bigg)
\nonumber\\
\nonumber\\
\mathcal{B}(r',\theta{}_s,z') &=& \atan \left(\frac{z'\left(r-x\cos\theta_s\right)\csc\theta_s}{x\mathbb{A}(r,x,\theta_s,z')} \right)\cos\theta_s-\atanh\left(\frac{z'}{\mathbb{A}(r,x,\theta_s,z')}\right)\sin\theta_s
\nonumber\\
\nonumber\\
\mathcal{C}(r',\theta{}',z_s) &=& -z_s\int \frac{\left(rx\cos\theta'-x^2-z_s^2\right)}{\mathbb{A}(r,x,\theta',z_s)\mathbb{B}(x,\theta',z_s)}\ud\theta' \nonumber\\
\nonumber\\
\mathcal{D}(r_s,\theta{}',z') &=& \frac{1}{4x^2}\left((r_s^2+x^2)\left(\ln\left\{\mathbb{A}(r_s,x,\theta',z')-z'\right\}-\ln\left\{\mathbb{A}(r_s,x,\theta',z')+z'\right\}\right)-2z'\mathbb{A}(r_s,x,\theta',z')\right)
\nonumber\\
\nonumber\\
\mathcal{E}(r_s,\theta{}',z') &=& -\frac{\sgn\left(\sin\theta'/2\right)}{r_{s}x\sqrt{(r_s+x)^2+z'^2}}\left(((r_s+x)^2+z'^2)E\left\{\mathbb{D}(\theta'),\mathbb{E}(r_s,x,z')\right\}\right.\nonumber\\
&&-\left.(r_s^2+x^2+z'^2)F\left\{\mathbb{D}(\theta'),\mathbb{E}(r_s,x,z')\right\}\right)
\nonumber\\
\nonumber\\
\mathcal{F}(r',\theta{}_s,z') &=& -\ln\left\{r-x\cos\theta_s+\mathbb{A}(r,x,\theta_s,z')\right\}
\nonumber\\
\nonumber\\
\mathcal{G}(r_s,\theta{}',z') &=& \frac{1}{r_s}\frac{1}{4x^2}\left((r_s^2-x^2)\left(\ln\left\{\mathbb{A}(r_s,x,\theta',z')-z'\right\}-\ln\left\{\mathbb{A}(r_s,x,\theta',z')+z'\right)\right\}-2z'\mathbb{A}(r_s,x,\theta',z')\right)
\nonumber\\
\nonumber\\
\mathcal{H}(r',\theta{}_s,z') &=& \int\frac{rx\left(1-\cos^2\theta_s\right)+z'^2\cos\theta_s}{\mathbb{A}(r,x,\theta_s,z')\mathbb{B}(x,\theta_s,z')}\ud z' \nonumber\\
\nonumber\\
\mathcal{I}(r_s,\theta{}',z') &=& -\frac{1}{r_s}\frac{\sgn\left(\sin\theta'/2\right)z'}{2x^2(r_s+x)\sqrt{(r_s+x)^2+z'^2}}\left((r_s-x)(r_s^2+x^2)\Pi\left\{\mathbb{D}(\theta'),\mathbb{C}(r_s,x),\mathbb{E}(r_s,x,z')\right\}\right.\nonumber\\
&&+\left.(r_s+x)\left(((r_s+x)^2+z'^2)E\left\{\mathbb{D}(\theta'),\mathbb{E}(r_s,x,z')\right\}-(2r_s^2+z'^2)F\left\{\mathbb{D}(\theta'),\mathbb{E}(r_s,x,z')\right\}\right)\right)
\nonumber\\
\nonumber\\
\mathcal{J}(r_s,\theta{}',z') &=& -\frac{1}{r_s}\frac{\sqrt{r_s^2-2r_sx\cos\theta'+x^2+z'^2}}{x}
\nonumber\\
\nonumber\\
\mathcal{K}(r',\theta{}',z_s) &=& \int\cos\theta'\ln\left(r-x\cos\theta'+\mathbb{A}(r,x,\theta',z_s)\right)\ud\theta'\nonumber\\
&&-\int\frac{\cos\theta'\left(2x^2r\cos^2\theta'-\cos\theta'(x^3-xz_s^2)-r(x^2+z_s^2)\right)}{\mathbb{A}(r,x,\theta',z_s)\mathbb{B}(x,\theta',z_s)}\ud\theta'\nonumber\\
&&+ \int\frac{x(rx\cos\theta'-x^2-z_s^2)}{\mathbb{A}(r,x,\theta',z_s)\mathbb{B}(x,\theta',z_s)}\ud\theta'
\nonumber\\
\nonumber\\
\mathcal{L}(r',\theta{}',z_s) &=& \frac{1}{2rx\sqrt{x^2+z_s^2}}\left(2r(x^2+z_s^2)\left(\atanh \mathbb{F}_+-\atanh \mathbb{F}_- \right)\right.\nonumber\\
&&+\left.\sqrt{x^2+z_s^2}\left(-(r^2+x^2+z_s^2)\ln\left\{r(r-x\cos\theta'+\mathbb{A}(r,x,\theta',z_s)\right\}\right.\right.\nonumber\\
&&+\left.\left.(r^2-2rx\cos\theta'+x^2+z_s^2)\ln\left\{r-x\cos\theta'+\mathbb{A}(r,x,\theta',z_s)\right\}\right.\right.\nonumber\\
&&-\left.\left.2r\mathbb{A}(r,x,\theta',z_s)-(r^2+x^2+z_s^2)\ln2+2rx\cos\theta'-(r^2+x^2+z_s^2)\right)\right) \nonumber
  \end{align}
\end{strip}

The integrals $\mathcal{C},\mathcal{H}$ and $\mathcal{K}$ cannot fully be evaluated analytically and have to partially be computed numerically. We implement this by using the QuadPack numerical integration library \cite{QuadPack}, specifically the \verb+qags+ function. The remaining integrals can be expressed analytically. This means that the following components of the tensor field can be evaluated analytically, $N_{xy}, N_{xz}, N_{yz}$ and $N_{zy}$. The remaining components have to be evaluated numerically, which are $N_{xx}, N_{yx}, N_{yy}, N_{zx}$ and $N_{zz}$.  Interestingly, as the tensor field is symmetric, as is the demagnetization tensor, it is curious that $N_{xz}$ and $N_{xy}$ have analytical expressions while $N_{zx}$ and $N_{yx}$ have to be evaluated numerically. However, when evaluated, the expressions have the same numerical value, and the tensor is thus symmetric.

\section{Verification}
The magnetic field as computed using the tensor field formulation is verified against a finite element method computation of the magnetic flux density from a cylindrical tile. The magnetic field is computed using the finite element framework Comsol. The equation solved in the FEM framework is the magnetic scalar potential equation
\begin{eqnarray}
-\boldsymbol{\nabla}{}\cdot{}(\mu{}_{0}\boldsymbol{\nabla}{}V_\mathrm{m})=-\boldsymbol{\nabla}{}\cdot{}(\mu{}_{0}\mathbf{M} )~.\label{Eq.Numerical_Magnetism}
\end{eqnarray}
The magnetic field is then calculated as $-\boldsymbol{\nabla}{}V_\mathrm{m} = \mathbf{H}$. For the FEM model, a highly refined finite element mesh is used to ensure a high precision in the computed field, and a sufficiently large volume is modelled to make the boundary conditions not influence the generated magnetic field.

\begin{figure}[!t]
  \centering
  \includegraphics[width=1\columnwidth]{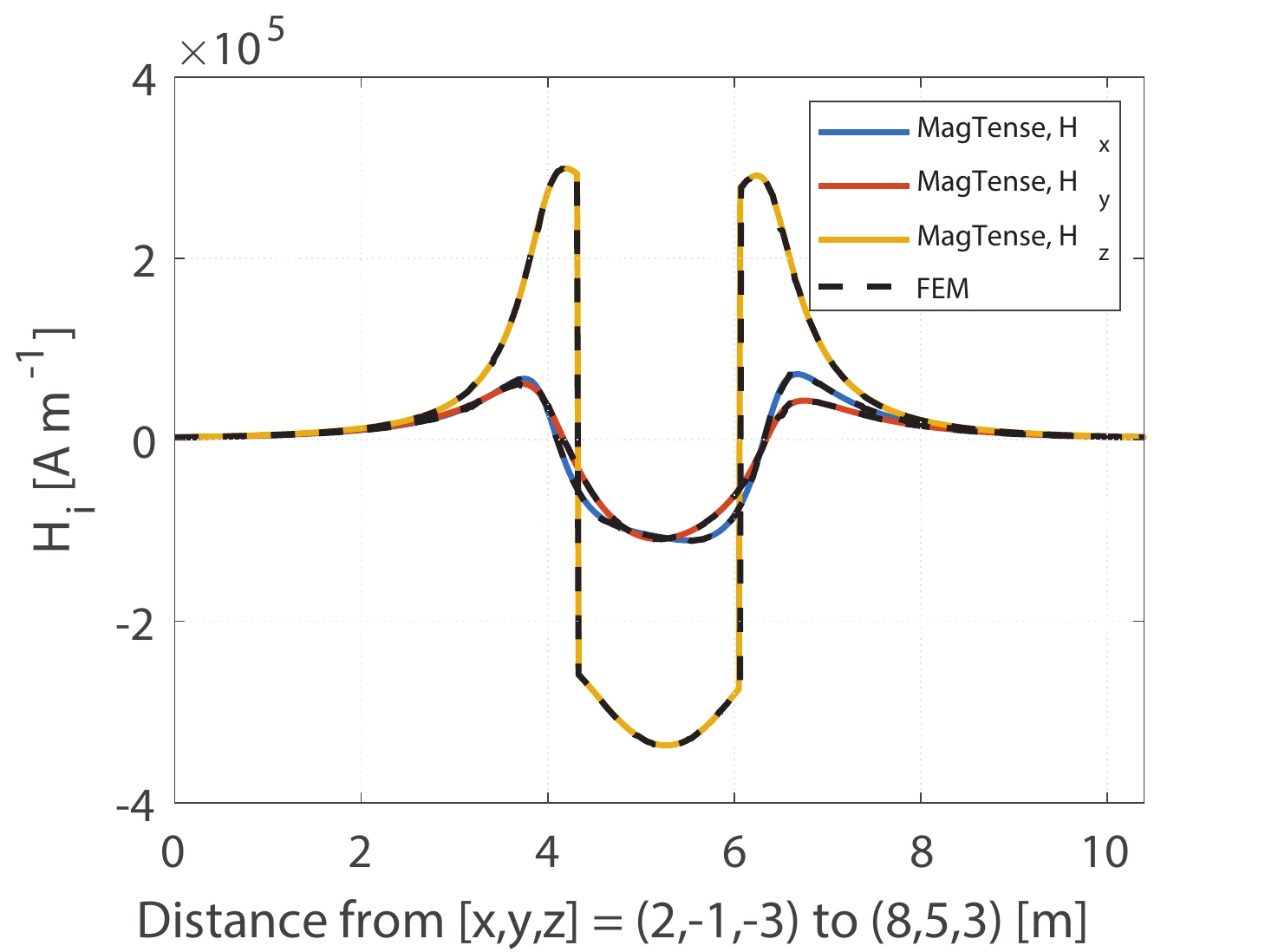}
  \caption{The components of the magnetic field along a line from the point $(x,y,z) = (2,-1,-3)$ to $(8,5,3)$ as calculated using the components of the tensor field given above in the MagTense framework and as calculated using the FEM framework Comsol, for the cylindrical tile geometry specified in the text.}
  \label{Fig_Compare_Comsol_example_1}
\end{figure}

In the first example we consider a cylindrical tile similar to the one shown in Fig. \ref{Fig_Cylinder_tile}. It has the following geometry: $r_1=4.3296$ mm, $r_2=6.4672$ mm, $\theta_1=0$, $\theta_2=\pi/4$, $z_1=-0.5$ mm and $z_2=0.5$ mm. This is similar to specifying the center of the tile with $(r_0,\theta_0,z_0)=(5.3984, \pi/8, 0)$ and tile dimensions of $(\Delta{}r,\Delta{}\theta,\Delta{}z)=(2.1376 \textrm{mm},\pi/8,1 \textrm{mm})$. The magnetization is specified by the vector \\$\mu_{0}\mathbf{M}=[0.6929,\; 0.6929,\; 0.6929]$ T. The components of the magnetic field is calculated along a line from the point $[x,y,z] = (2,-1,-3)$ to $(8,5,3)$, thus passing almost through the center of the tile and is shown in Fig. \ref{Fig_Compare_Comsol_example_1}. As can be seen from the figure, there is an excellent agreement between the model framework presented here and the FEM model. This example is part of the verification examples for MagTense and is available online \cite{MagTense}.

\begin{figure}[!t]
  \centering
  \includegraphics[width=1\columnwidth]{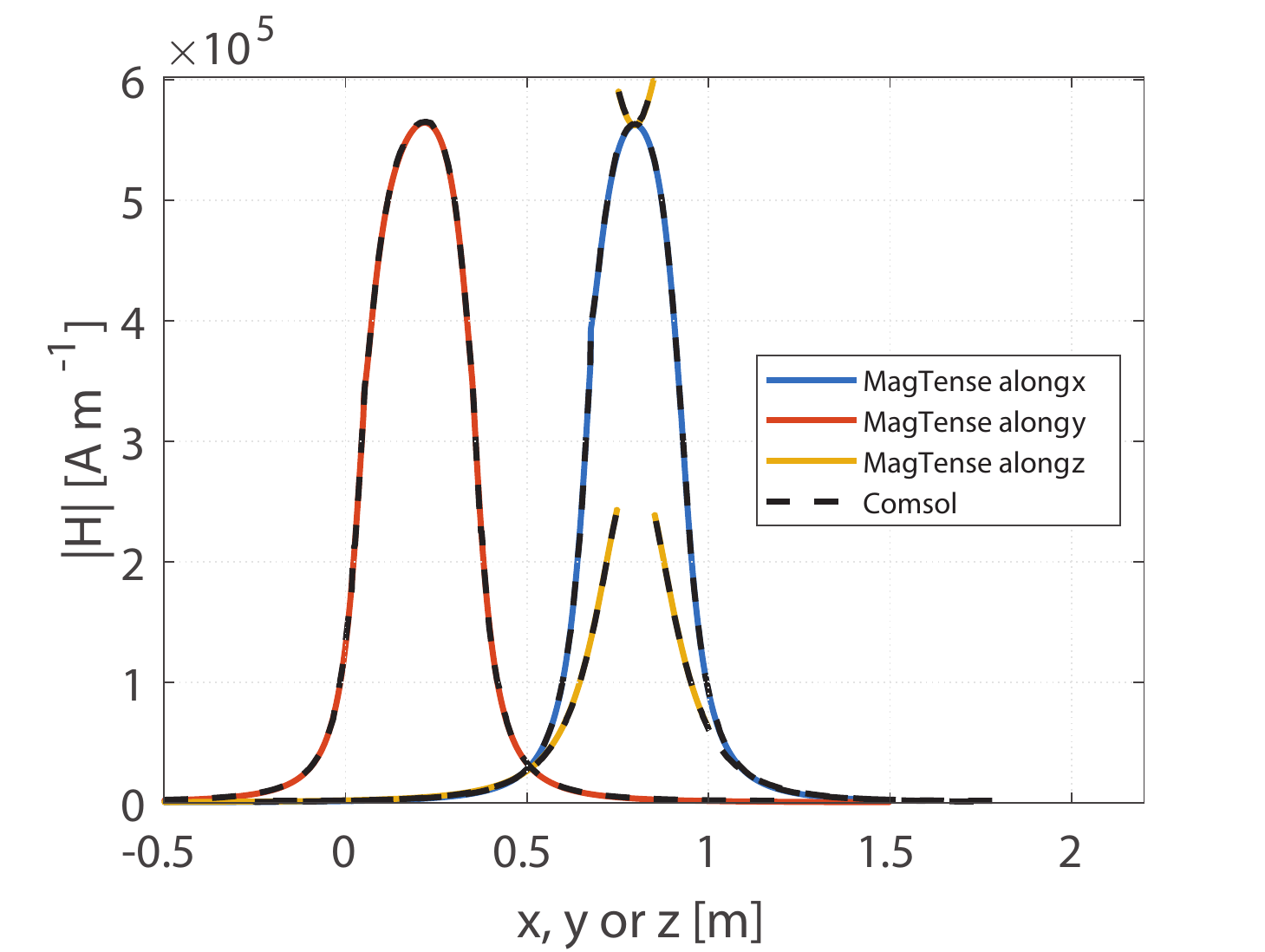}
  \caption{The norm of the magnetic field as calculated using the components of the tensor field given above in MagTense and as calculated using the FEM framework Comsol, for the cylindrical tile geometry specified in the text along each of the Cartesian axes through the center of the tile.}
  \label{Fig_Compare_Comsol_example_2}
\end{figure}

As a second example, we consider a cylindrical tile with the following geometry: $r_1=150$ mm, $r_2=450$ mm, $\theta_1=67.5^\circ$, $\theta_2=112.5^\circ$, $z_1=750$ mm, $z_2=850$ mm, with the cylinder central axis offset to $x_c=800$ mm, $y_c=-100$ mm. The magnetization is specified by the vector \\$\mu_{0}\mathbf{M}=[0.424,\; 0.424,\; 1.04]$ T, which corresponds to a direction specified by the spherical azimuthal angle $\phi=\pi/4$ and the spherical polar angle $\theta=\pi/6$ and a magnitude of $M=1.2/\mu_{0}$. The norm of the magnetic field is calculated along each of the Cartesian axes centered on the tile and is shown in Fig. \ref{Fig_Compare_Comsol_example_2}. As can be seen from the figure, there is again an excellent agreement between the model framework presented here and the FEM model. The individual components of the field, which are not shown in the figure due to brevity, show an equally excellent agreement. This example is part of the verification examples for MagTense and is available online \cite{MagTense}.

Regarding the computational speed of calculating the magnetic field, in the FEM model this crucially depends on the number of elements used, as do the precision. For example one the FEM solution has a fine mesh with $2.2*10^6$ tetrahedral elements. On an Intel Xeon W-3235 with 128 GB of memory the computation time for this FEM model was 47 seconds (disregarding the meshing time, which was about the same time). For MagTense the computation time was 0.126 seconds. For example two, an almost equal mesh with also $2.2*10^6$ tetrahedral element was used. Here the computational time was 43 seconds, while the MagTense computation time was 0.39 seconds, as three time as many points are evaluated in example two compared to example one. The computational speed of MagTense is so fast because in MagTense the field is only evaluated in a set of specified points, whereas it in a FEM model has to be evaluated in the entire simulation volume.

\section{Conclusion}
We have calculated the magnetic flux density components formulated as a tensor field for a cylindrical tile geometry. The components of the tensor field involved a number of integrals, which meant that not all components could be evaluated analytically. In the end, five components $N_{xx},N_{yx},N_{yy},N_{zx}$ and $N_{zz}$ have to be evaluated numerically, while $N_{xy},N_{xz},N_{yz}$ and $N_{zy}$ could be evaluated purely analytically. As the tensor field is symmetric, only six of these components needs to be evaluated in actual computations. There was an excellent agreement between the magnetic flux density of a cylindrical tile calculated using the tensor field and the flux density computed using an finite element approach.

\section*{Acknowledgements}
This work was financed partly by the Energy Technology Development and Demonstration Program (EUDP) under the Danish Energy Agency, project no. 64016-0058, partly by the Danish Research Council for Independent Research, Technology and Production Sciences projects no. 7017-00034 and 8022-00038 and partly by the Poul Due Jensen Foundation project on Browns paradox in permanent magnets, project no. 2018-016.

\appendix
\section{Singularities in the solutions}
Some of the above given integrals have singularities for certain values of the variables. This puts limitations on the possible coordinates at which the tensor field can be evaluated. It is obvious that the following cannot be violated:
\begin{eqnarray}
r&>&0\\
x&\neq&0\\
\mathbb{A}(r,x,\theta',z)&>&0\\
\mathbb{B}(x,\theta',z)&\neq&0\\
r-x\cos\theta'+\mathbb{A}(r,x,\theta',z)&>&0\label{ineq_r_x_A}\\
\mathbb{A}(r,x,\theta,z)-z&>&0\label{ineq_A_min_z}\\
\mathbb{A}(r,x,\theta,z)+z&>&0\label{ineq_A_plus_z}\\
\mathbb{F}_\pm&\neq& \pm 1\nonumber\\
\Rightarrow \mathbb{A}(r,x,\theta,z)\pm r\neq \pm\sqrt{x^2+z^2},\label{eq_F_pm}
\end{eqnarray}
where $n$ is an integer.

\subsubsection{The $\mathbb{A}$ function}
It is a requirement that $\mathbb{A}$ must be real and greater than zero and thus $r^2-2xr\cos\theta+x^2+z^2>0$. In the following, we will consider the three variables, $x$, $r$ and $\theta$ for any real value of $z$. First, we note that $\mathbb{A}$ is symmetric in $x$ and $r$ so the results derived from one of these variables will apply equally to the other.

We consider the above given inequality in the limit where it is an equality and solve for $x$:
\begin{eqnarray}
x=r\cos\theta\pm\sqrt{r^2\cos^2\theta-(r^2+z^2)}.
\end{eqnarray}
We see immediately that $x$ can only be real valued when $z=0\ \wedge\ \theta=n\pi$ for $n$ being any integer including zero and conclude that the inequality holds for any value of $x$ and therefore also $r$ with this requirement.

Considering $\theta$, we get that $2xr\cos\theta\le r^2+x^2+z^2=(r+x)^2-2xr+z^2$, which is to say that for $\cos\theta\le0$ the inequality always holds. However, in the case where $z=0\ \wedge\ \cos\theta>0$ we get:
\begin{eqnarray}
\cos\theta\le \frac{r^2+x^2}{2xr}=\frac{1}{2}\left(\frac{r}{x}+\frac{x}{r}\right).
\end{eqnarray}
This expression becomes an equality in the limit where $\cos\theta=1\ \wedge\ r=x$, which means that $\mathbb{A}$ is real for all values of $\theta$ also when $z\neq 0$. In conclusion, we have the requirement that if $z=0$ then the solution is ill determined for values $\theta=n\pi$.

\subsubsection{The $\mathbb{B}$ function}
The function $\mathbb{B}=x^2(\cos^2\theta-1)-z^2$ cannot be zero. By solving for $x,\ z$ and then $\theta$, respectively, we can see that only when $z=0\ \wedge\ \theta=n\pi$ for $n$ being an integer or zero will $\mathbb{B}$ be zero.

\subsubsection{Inequality \ref{ineq_r_x_A}}
The inequality in \ref{ineq_r_x_A} may be written as
\begin{eqnarray}
r^2+x^2\cos^2\theta-2xr\cos\theta&<&r^2+x^2+z^2-2xr\cos\theta\nonumber\\
\Rightarrow \cos^2\theta-1&<&\frac{z^2}{x^2}.\label{ineq_constr_r_x_A}
\end{eqnarray}
We already have the requirement that $x\neq0$ and can thus see from \ref{ineq_constr_r_x_A} that the condition in Eq. \ref{ineq_r_x_A} is violated only when $z=0\ \wedge\ \theta=n\pi$.

\subsubsection{Inequality \ref{ineq_A_min_z}}
The inequality in \ref{ineq_A_min_z} has the, in $x$ and $r$, symmetric solution in the limit of $\mathbb{A}-z=0$:
\begin{equation}
r=x\ \wedge\ \theta=n\pi\ \wedge\ x>0,
\end{equation}
for any integer (including zero) value of $n$. We note in passing that Eq. \ref{ineq_A_plus_z} is violated only when $z=0\ \wedge\ \theta=n\pi$.

\subsubsection{Equality \ref{eq_F_pm}}
The equality in Eq. \ref{eq_F_pm} solved for the violating condition, i.e. $A\pm r=\pm\sqrt{x^2+z^2}$, has the solution
\begin{equation}
\cos^2\theta-1=\frac{z^2}{x^2},
\end{equation}
which is to say that, again, we have the violating condition $z=0\ \wedge\ \theta=n\pi$.

\subsubsection{Summary of constraints}
In summary, we have the following conditions where the solution presented here is not valid:
\begin{eqnarray}
\theta=n\pi\ \wedge\ (z=0\ \vee\ r=x\ \wedge\ x>0)\ \vee\ x=0\ \vee\ r=0.
\end{eqnarray}


\end{document}